\documentclass[twocolumn,showpacs,amsmath,amssymb,groupaddresses,superscriptaddress]{revtex4}
\usepackage{times}
\usepackage{latexsym}
\usepackage{graphicx}
\usepackage{amsmath,amssymb}
\usepackage{verbatim,bbm}

\begin{document}

\title{Continuous-variable quantum enigma machines for long-distance key distribution}

\author{Cosmo Lupo}
\address{Research Laboratory of Electronics, Massachusetts Institute of Technology, Cambridge, MA 02139, USA}

\author{Seth Lloyd}
\affiliation{Department of Mechanical Engineering, Massachusetts Institute of Technology, Cambridge, MA 02139, USA}
\affiliation{Research Laboratory of Electronics, Massachusetts Institute of Technology, Cambridge, MA 02139, USA}

\begin{abstract}
Quantum physics allows for unconditionally secure communication through insecure communication channels.
The achievable rates of quantum-secured communication are fundamentally limited by the laws of quantum 
physics and in particular by the properties of entanglement. 
For a lossy communication line, this implies that the secret-key generation 
rate vanishes at least exponentially with the communication distance.
We show that this fundamental limitation can be violated in a realistic scenario where 
the eavesdropper can store quantum information for only a finite, yet arbitrarily long, time.
We consider communication through a lossy bononic channel (modeling linear loss in optical fibers)
and we show that it is in principle possible to achieve a constant rate
of key generation of one bit per optical mode over arbitrarily long communication distances.
\end{abstract}


\maketitle

\section{Introduction}

Quantum key distribution (QKD) promises unconditional secure communication through insecure
communication channels \cite{QKD}. 
In real world implementations of QKD, however, the achievable secret-key rates are still 
relatively low compared with standard telecommunication rates.
The rates of secret-key generation are not only constrained by experimental imperfections,
which can be amended in principle, but are also limited by the fundamental features of quantum physics.
As recently shown in \cite{TGW}, the entanglement between the two
ends of the communication channel ultimately bounds the 
maximum rate of secret-key generation:
\begin{equation}\label{TGW_1}
R \leq E_\mathrm{sq}(\mathcal{N}) \, ,
\end{equation}
where $E_\mathrm{sq}(\mathcal{N})$ is an entropic quantity called the squashed entanglement 
of the channel \cite{squash}, which is function of the quantum communication channel $\mathcal{N}$
linking the legitimate sender Alice to the legitimate receiver Bob.

In this paper we consider the case where the communication channel $\mathcal{N}$
is a lossy (and noisy) bosonic channel. This means that information is encoded in a
collection of bosonic modes whose corresponding canonical operators are denoted
$a_j$, $a_j^\dag$ and satisfy the commutation relations $[ a_{j'} , a_j^\dag ] = \delta_{jj'}$.
In the Heisenberg picture the quantum channel maps the canonical operators $a_j$, $a_j^\dag$
to $a_j \to \sqrt{\eta} \, a_j + \sqrt{1-\eta} \, v_j$, 
$a_j^\dag \to \sqrt{\eta} \, a_j^\dag + \sqrt{1-\eta} \, v_j^\dag$, where $\eta \in [0,1]$ is
the attenuation factor (also called transmissivity) and $v_j$, $v_j^\dag$ are the
canonical ladder operators of an environment bosonic mode.
The lossy channel is obtained if the environment mode is initially in the vacuum state,
while the lossy and noisy channel corresponds to the environment mode being in a thermal
state with $N_T$ mean photon number.
These channels attenuate the input power by a 
factor $\eta$ and model the ubiquitous processes of linear absorption and scattering
of light.

When applied to the case of the lossy bosonic channel, the squashed entanglement bound in (\ref{TGW_1}) yields \cite{TGW}: 
\begin{equation}\label{TGW_2}
R \leq \log{\left( \frac{1+\eta}{1-\eta}\right) } \, ,
\end{equation}
where the rate is measured in bits (throughout this paper $\log \equiv \log_2$)
per bosonic mode (given the bandwidth of the channel, this can be easily translated
in bits per second).
For both free space and fiber optics communication, the attenuation factor $\eta = e^{-\ell/\ell_0}$
scales exponentially with the distance $\ell$ between sender and receiver, where the characteristic length
$\ell_0$ depends on experimental conditions.
For long distances, $R \leq 2 \,\eta = 2 \, e^{-\ell/\ell_0}$, and 
the key rate decays at least exponentially with increasing communication distance.
This result marks a striking difference between quantum-secured communication and (insecure) 
classical communication. In the latter case, one can in principle achieve a finite 
communication rate over arbitrarily long distances, just by sufficiently 
increasing the signal power \cite{Gconj0}.
Unfortunately, this is not the case for quantum communication where a fundamental
rate-distance tradeoff exists, requiring the use of quantum repeaters to perform QKD 
on long distances.

It is thus clear that to go around the fundamental rate-distance tradeoff in
(\ref{TGW_2}) one should renounce unconditionally security.
Here we discuss QKD conditioned on the assumption 
that technological limitations allow an eavesdropper Eve to store quantum information reliably 
only for a known and finite -- but otherwise arbitrarily long -- time.
Such an eavesdropper may also have unlimited computational power, including 
a quantum computer.
Indeed, any physical realization of a quantum memory can reliably store 
quantum information only for a time of the order of its coherence time.
We stress that we do not require the legitimate receiver to have better quantum 
storage technologies than the eavesdropper. 
As will be shown, the legitimate parties could have a much shorter memory time than
the eavesdropper and the communication will still be secure.

%---

\section{Security definitions}

According to the state of the art, one requires a quantum cryptography protocol to be
unconditionally and composably secure.
Unconditional security means that one does not rely on unproven statements (e.g, about the complexity
of factorizing large numbers, or in general about the computational power of the eavesdropper).
Composable security means that the given protocol is secure also when used as a subroutine
within an overarching protocol \cite{compo}.

Suppose that a given communication protocol aims at establishing
a secret message described as a random variable $X$. The information about $X$ in the 
hands of the eavesdropper Eve is described, without loss of generality, by a bipartite quantum state 
of the form 
\begin{equation}
\rho_{XE} = \sum_x p_X(x) \, |x\rangle \langle x| \otimes \rho_E(x) \, .
\end{equation}
Ideally, one would like Eve's state to be completely uncorrelated with
the message $X$, that is, $\rho_{XE} = \rho_X \otimes \rho_E$ \cite{NOTA_prod}.
To quantify the deviation from such an ideal setting one considers the trace distance \cite{NOTA_tdist}
\begin{equation}
D(\rho_{XE},\rho_X \otimes \rho_E) := \frac{1}{2} \, \| \rho_{XE} - \rho_X \otimes \rho_E \|_1 \, .
\end{equation}
Therefore, the security of the communication protocol is assessed by the condition
\begin{equation}
D(\rho_{XE},\rho_X \otimes \rho_E) \leq \epsilon \, ,
\end{equation}
which implies that the state $\rho_{XE}$ is indistinguishable, up to a probability
smaller than $\epsilon$, from the state $\rho_{X} \otimes \rho_E$, that is, 
the given communication protocol is secure up to a probability smaller than $\epsilon$ \cite{Renner}.
As a matter of fact this criterion guarantees unconditional and composable security \cite{Renner}.

%--- 

In this paper we renounce unconditional security and seek security conditioned on the
fact that the eavesdropper can store quantum information only for a finite and known time $\tau$.
This means that Eve is forced to make a measurement within a time $\tau$ after obtaining the
quantum state.
Suppose that Eve has made a measurement $\Lambda$ described by the POVM (positive operator valued measurement) 
elements $\{ \Lambda_y \}_y$ \cite{NOTA_POVM}.
After the measurement has been made, the state has `collapsed' to 
\begin{align}
\rho'_{XE} & = \sum_y \mathrm{Tr}_E( \rho_{XE} \, \mathbb{I} \otimes \Lambda_y ) \, |y\rangle_E \langle y | \\
& = \sum_{x,y} p_X(x) \, \mathrm{Tr} \left( \rho_{E}(x) \, \Lambda_y \right) \, |x\rangle \langle x | \otimes  |y\rangle_E \langle y | \, .
\end{align}
Since $\rho'_{XE}$ is diagonal in the basis $\{ |x\rangle \otimes |y\rangle \}$, we have
\begin{align}
D(\rho'_{XE} , \rho'_X \otimes \rho'_E) & = \sum_{x,y} \left| p_{XY}(x,y) - p_X(x) p_Y(y) \right| \\
& =: D(p_{XY}, p_X p_Y) \, ,
\end{align}
where $p_{XY}(x,y) = p_X(x) p_{Y|x}(y)$ with $p_{Y|x}(y) = \mathrm{Tr} \left( \rho_{E}(x) \, \Lambda_y \right)$
and $p_Y(y) = \sum_x p_X(x) p_{Y|x}(y)$, that is, the trace distance equals 
the distance between classical probabilities.
Finally, optimizing over Eve's choice of her measurement, we obtain the following
security condition:
\begin{equation}\label{cond1}
\sup_{\Lambda} D(p_{XY}, p_X p_Y) \leq \epsilon \, .
\end{equation}

In this paper, instead of working directly with condition (\ref{cond1}), we require
\begin{equation}
I_\mathrm{acc}(X;E)_{\rho} \leq \epsilon' \, ,
\end{equation}
where $I_\mathrm{acc}(X;E)_{\rho}$ denotes the accessible information of Eve about $X$
given the state $\rho_{XE}$ \cite{NOTA_acc}.
The latter implies condition (\ref{cond1}), for $\epsilon = \sqrt{ 2\ln{(2)} \, \epsilon' }$, 
via Pinsker inequality \cite{NOTA_Pinsker}
\begin{equation}
\max_\Lambda D(p_{XY}, p_X p_Y) \leq \sqrt{ 2\ln{(2)} \, I_\mathrm{acc}(X;E)_{\rho} } \, .
\end{equation}

%---

It is worth recalling that accessible information was used as a security quantifier 
during the first years of quantum cryptography, since it was found that a security criterion
based on the accessible information does not in general guarantee composable security in an unconditional
manner \cite{Renner}.
Here instead we have shown that composability holds under condition (\ref{cond1}) 
if we give up full unconditional security and seek security under the assumption that the 
eavesdropper can store quantum information only for a finite and known time --- 
i.e, she has a quantum memory with limited storage time.

%---

\section{Summary of the results}

We present two novel key-generation protocols for continuous-variable quantum optical communication
through a lossy bosonic channel with transmissivity $\eta$, modeling linear attenuation 
and scattering. 
These protocols are composably secure under the condition that Eve's as a quantum memory
with finite, and known, but otherwise arbitrarily long, storage time.

\begin{figure}
\centering
\includegraphics[width=0.35\textwidth]{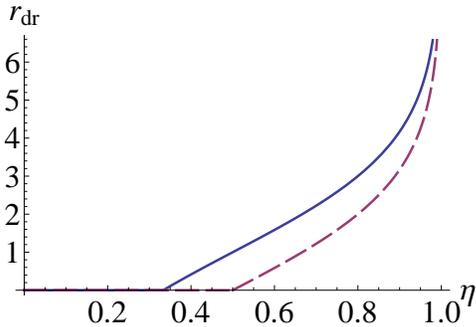}
\caption{
Achievable key rate for the pure loss channel ($N_T=0$) vs the channel transmissivity $\eta$, in bits per mode,
for direct reconciliation protocols. 
Blue solid line: Achievable locked-key rate as given by the expression in (\ref{rate-dr-infty}).
Red dashed line: Maximum fully unconditional secret-key rate,
given by the expression $\max\{ 0 , \log{\left( \frac{\eta}{1-\eta} \right)} \}$ \cite{Wolf}.}
\label{fig:direct}
\end{figure}

The first protocol is a direct-reconciliation protocol (in which we allow information reconciliation
by forward public communication from the sender Alice to the receiver Bob). 
We obtain a simple formula for the asymptotic key rate (see Fig.\ \ref{fig:direct}):
\begin{equation}\label{rate-dr-infty}
r_\mathrm{dr} = 1 + \log{\left(\frac{\eta}{1-\eta}\right)} \, .
\end{equation}
This protocol can generate a nonzero key rate for any $\eta > 1/3$. 
By comparison, the maximum unconditionally secure key rate from direct reconciliation
is given by the quantum capacity formula $\log{\left( \frac{\eta}{1-\eta} \right)}$ \cite{Wolf}
and is positive only for $\eta > 1/2$ \cite{sext}.

\begin{figure}
\centering
\includegraphics[width=0.35\textwidth]{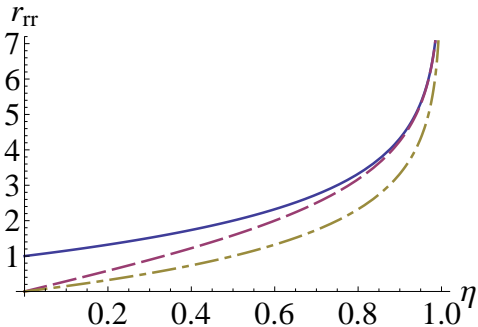}
\caption{
Achievable key rate for the pure loss channel ($N_T=0$) vs the channel transmissivity $\eta$, in bits per mode,
for reverse reconciliation protocols. 
Blue solid line: Achievable locked-key rate as from the expression in (\ref{lkrr}).
Red dashed line: Upper bound for the secret-key rate (assisted by two-way public communication),
given by the expression in (\ref{TGW_2}).
Yellow dash-dotted line: Achievable asymptotic secret-key rate 
according to the standard security definition as given by the reverse
coherence information $\log{\left( \frac{1}{1-\eta} \right)}$ \cite{Pirs}.}
\label{fig:reverse}
\end{figure}

The second protocol is a reverse-reconciliation protocol (we allow information reconciliation
by backward public communication from Bob to Alice). In this setting we show that Alice and
Bob can in principle generate key at an asymptotic rate of more than $1$ bit per bosonic mode sent
through the channel. This is true for any nonzero value of the transmissivity $\eta$, provided sufficient
input energy is provided --- hence reproducing the feature of insecure classical communication in a
quantum-secured communication framework. The achievable asymptotic key rate is (see Fig.\ \ref{fig:reverse})
\begin{equation}\label{lkrr}
r_\mathrm{rr} = 1 + \log{\left(\frac{1}{1-\eta}\right)} \, .
\end{equation}
By comparison, the maximum fully unconditional key rate is upper bounded by the expression in
(\ref{TGW_2}) and vanishes as $2\eta$ for small values of $\eta$.

\begin{figure}
\centering
\includegraphics[width=0.35\textwidth]{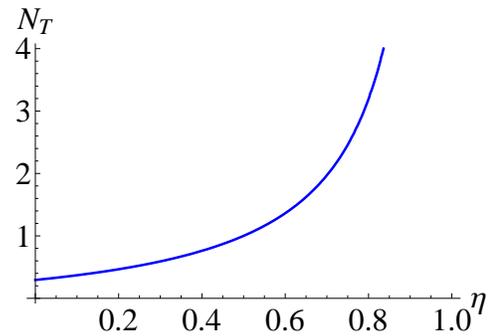}
\caption{
Tolerable excess noise $N_T$ vs the transmissivity $\eta$
for the reverse-reconciliation quantum data locking protocol, from Eq.\ (\ref{thermal}).
The asymptotic locked-key generation rate is nonzero for values of
$(\eta,N_T)$ below the curve.}
\label{fig:eccess}
\end{figure}

We also consider the case of lossy and noisy bosonic channel, which models the presence of experimental imperfection
or a thermal-like background with $N_T$ mean photons per mode. 
The lossy and noisy channel is also used to model an `active attack' from the eavesdropper, who
injects noise in the channel. 
In this case we obtain an asymptotic rate equal to
\begin{equation}\label{thermal}
r_\mathrm{rr} = 1 + \log{\left( \frac{1}{1-\eta} \right) } - g(N_T) \, ,
\end{equation}
which is nonzero at arbitrary distances provided $N_T \lesssim 0.3$ (see Fig.\ \ref{fig:eccess})

These protocols are instances of quantum data locking protocols (see Sec.\ \ref{sec:QDL}).
We henceforth call {\it locked key} a key which is generated by a quantum data locking
protocol, just to remind us that this key is not unconditionally secure, but secure conditioned on
the assumption of finite memory storage time.

\section{Comparison with other models}
It is known that high rates of secret-key generation can be attained
against an eavesdropper endowed with an imperfect quantum memory, as for
example in the Bounded Storage Model, where Eve can store only a constrained
number of qubits (see e.g.\ \cite{BSM}).
Even under bounded storage, no known protocol attains a constant rate as a function of distance.
Elsewhere we have shown that quantum data locking allows for a substantial 
enhancement of the key rate \cite{PRL,NJP}.
Here we show for the first time that such an assumption allows us to
generate key at a constant rate across virtually any distance.
It is an open question whether the quantum data locking could be applied
in the bounded storage model to attain rates of key generation independent
on the distance.

Our results must be compared with the bounds on the optimal secret-key
rate obtained requiring fully unconditional security.
In the asymptotic setting, the security is usually quantified by the 
quantum mutual information (see e.g. \cite{PrivateD}). 
The gain in key generation rate that we achieve
follows from the existence of a large gap between the quantum mutual information and
the accessible information of the adversary. This gap is well known in quantum
information theory: it is the {\it quantum discord} \cite{QD}, 
which quantifies the quantum correlations that the adversary cannot access by local measurements 
on her share of the quantum system. 

%---

\section{Quantum data locking and quantum enigma machines}\label{sec:QDL}

In a typical quantum data locking protocol \cite{QDL,CMP,Buhrman,Leung}, the two legitimate parties,
say Alice and Bob, publicly agree on a set of $MK$ quantum codewords. 
They then use a preshared secret key of $\log{K}$ bits, labeled by $s=1,2,\dots,K$, to secretly agree on a set of $M$ 
(equally probable) codewords, labeled by $x=1,2,\dots,M$, used to
encode $\log{M}$ bits of classical information.
These quantum codewords are sent through $n$ uses of a quantum channel from Alice to Bob.
Suppose an eavesdropper Eve tampers with the communication line and obtains one of the states
$\rho^n_E(x,s)$.
The correlations between Eve's quantum system and the input message $x$ are described
by the state
\begin{equation}
\rho^n_{XE} = \frac{1}{M} \sum_{x=1}^M |x\rangle\langle x| \otimes \frac{1}{K}\sum_{s=1}^K \rho^n_E(x,s) \, ,
\end{equation}
where $\{ |x\rangle \}_{x=1,\dots,M}$ is an orthonormal basis for an auxiliary quantum
system encoding the messages $x$ --- notice that the summation over $s$ comes from the fact that
Eve does not know the value of the secret key.
One can prove that, if the states $\rho^n_E(x,s)$ have a suitable form and for $K$ large enough, Eve
can only obtain a negligible amount of the classical information
--- as quantified by the accessible information --- carried by the label $x$.

In the most powerful quantum data locking schemes known up to now, a constant-size 
preshared secret seed of about $\log{K} = \log{1/\epsilon}$ bits 
allows Alice and Bob to encrypt $\log{M}$ bits (with $M$ arbitrarily large), with the guarantee 
that Eve's accessible information is of the order of $\epsilon \log{M}$ bits \cite{Fawzi,Dupuis,phase}.

It is worth remarking that quantum data locking provides a strongest violations of 
classical information theory in the quantum setting. Indeed, according to a famous theorem
of Shannon's, which assesses the security of one-time pad encryption, 
to encrypt $m$ bits of classical information Alice and Bob need at least $m$ bits
of preshared secret key \cite{Shannon}. 
Quantum data locking violates this Shannon's result by an exponential amount.

A quantum data locking protocol can be seen as a quantum counterpart of the twentieth century
Enigma machine \cite{QEM}. Following \cite{QEM,PRX} we call `quantum enigma machine' an 
optical cipher that harnesses the quantum data locking effect.

\subsection{Quantum bootstrapping}\label{sec:boot}

The first works on quantum data locking only considered the ideal case of a noiseless communication scenario.
Only recently the quantum data locking effect has been considered in a noisy setting \cite{QEM,PRX,AW} (see also \cite{Boixo}).
Here we combine quantum data locking with a key-recycling technique
that has been successfully applied to quantum data locking in a noisy communication scenario \cite{PRL,QDL,Entropy}.

We assume that eavesdropper Eve and the legitimate receiver can store quantum information 
for a time $\tau_E$ and $\tau_B$, respectively.

Suppose then that Alice and Bob, using the quantum channel $n$ times, run a
quantum data locking protocol to communicate $\log{M} = n \chi$ bits of classical information, 
and consume $\log{K} = n k$ bits of preshared secret key.
Bob may need to perform a collective measurement over $n$ quantum systems in order to decode.
Since, as from our assumption, Bob's quantum memory can store quantum information 
only for times shorter than $\tau_B$, this requires that the $n$ quantum signals should be sent
within this time interval (this is always possible for $\tau_B$ large enough or by
increasing the repetition rate).

On the other hand, if Eve has a quantum memory with finite coherence time $\tau_E$, this implies that
she is forced to measure within a time $\tau_E$ after receiving the signals, 
otherwise her memory will decohere anyway.
Therefore, what the legitimate parties Alice and Bob can do is to wait for a time longer than $\tau_E$
before sending more information through the channel. 
After waiting such a time, Alice and Bob can safely recycle part of the obtained key
as a fresh key to run another round of quantum data locking.

Thus, for $\chi > k$, Alice and Bob can recycle part of the newly
established key and use it as a seed for another round of quantum data locking.
By repeating this procedure many times they will asymptotically obtain a overall locked-key rate of 
$r = \chi - k$ bits per channel use, with a negligible amount of initially shared secret key.

While $r = \chi - k$ is the rate of bits per channel use, one could expect a lower 
rate in terms of bits per second, due to the waiting times between quantum data locking subroutines.
There is a simple strategy to solve this problem: Alice and Bob can use the dead times to
run two (or more) independent quantum data locking protocols. 
In this way they can in principle
achieve a rate of bits per second as high as $r \nu = (\chi - k) \nu$, where $\nu$ is the number of channel
uses per second.
Notice that this holds for any value of $\tau_E$, as long as it is known to Alice and Bob, 
and independently of $\tau_B$ (for instance we can take $\tau_B = \tau_E$ or even $\tau_B < \tau_E$).

%---

\section{The direct reconciliation protocol}\label{sec:dir}

\begin{figure}
\centering
\includegraphics[width=0.4\textwidth]{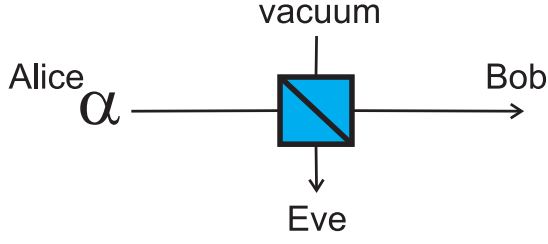}
\caption{The lossy bosonic channel can be modeled as a beam-splitter with
transmissivity $\eta$ and the environment mode initially in the vacuum state.
In the direct reconciliation protocol, Alice sends coherent state down the channel.}
\label{fig:ddirect}
\end{figure}

Alice prepares multimode coherent states that encode
both the input message $x \in \{ 1,\dots, M\}$ and the value
of the secret seeds $s \in \{ 1, \dots, K\}$ she shares with Bob. 
The encoding is by a random code (whose codebook is public) that assigns to each pair $(x,s)$
an $n$-mode coherent states
\begin{equation}
|\alpha^n(x,s)\rangle = \bigotimes_{j=1}^n |\alpha_j(x,s)\rangle \, ,
\end{equation}
where $\alpha_j(x,s)$ is the amplitude of the coherent state of the $j$-th bosonic mode
sent through the channel.
This is schematically depicted in Fig.\ \ref{fig:ddirect}, where the lossty channel is represented
as a beam-plitter.
To construct the random code, the amplitudes of the coherent states
are independently drawn from a circularly symmetric Gaussian distribution,
denoted $G_{(0,N)}$, with zero mean and mean photon number $\int d^2\alpha \, |\alpha|^2 \, G_{(0,N)} = N$.

The receiver Bob obtains the attenuated coherent states
\begin{equation}
|\sqrt{\eta} \, \alpha^n(x,s)\rangle = \bigotimes_{j=1}^n |\sqrt{\eta} \, \alpha_j(x,s)\rangle \, .
\end{equation}
The goal of Bob, who knows the value of $s$, is to decode $x$. 
It is known that he can do that (with asymptotically negligible error)
with an asymptotic bit-rate for $x$ given by \cite{Gconj0}
\begin{equation}
\chi_\mathrm{dr} := \lim_{n\to\infty} \frac{\log{M}}{n} = g(\eta N) \, ,
\end{equation}
where
\begin{equation}
g(N) = (N+1)\log{(N+1)} - N \log{N} \, .
\end{equation}

To guarantee the security of the communication protocol, we have to bound Eve's
accessible information.
For any $x$ and $s$, Eve obtains the attenuated coherent states
\begin{equation}
|\sqrt{1-\eta} \, \alpha^n(x,s)\rangle = \bigotimes_{j=1}^n |\sqrt{1-\eta} \, \alpha_j(x,s)\rangle \, .
\end{equation}
We can show (see Sec.\ \ref{sec:proofs}) that Eve's accessible information about Alice's input message $x$
is negligibly small, provided Alice and Bob 
initially share enough bits of secret key.
For $N$ large enough, and asymptotically in $n$, this is achieved for
\begin{equation}
k_\mathrm{dr} := \lim_{n\to\infty} \frac{ \log{K} }{n} = 2 g[(1-\eta)N] - g[2(1-\eta)N] \, .
\end{equation} 

Applying the key-bootstrapping routine (see Sec.\ \ref{sec:boot}), 
this yields a net asymptotic locked-key generation rate of 
\begin{equation}
r_\mathrm{dr} = \chi_\mathrm{dr} - k_\mathrm{dr} 
= g(\eta N) - 2 g[(1-\eta)N] + g[2(1-\eta)N] \, ,
\end{equation}
which in the limit of $N \to \infty$ becomes
\begin{equation}
r_\mathrm{dr} = 1 + \log{ \left( \frac{\eta}{1-\eta} \right) } \, .
\end{equation}

%---

\section{The reverse reconciliation protocol}\label{sec:rev}

\begin{figure}
\centering
\includegraphics[width=0.4\textwidth]{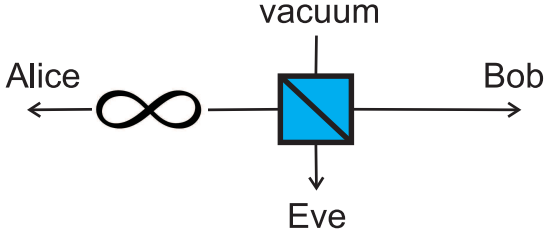}
\caption{The lossy bosonic channel can be modeled as a beam-splitter with
transmissivity $\eta$ and the environment mode initially in the vacuum state.
In the first phase of the reverse reconciliation protocol, Alice sends one mode of a two-mode
entangled state (denoted by the symbol `$\infty$') down the channel.}
\label{fig:dreverse}
\end{figure}

In the first phase of the protocol Alice prepares $n$ instances of a two-mode squeezed vacuum state, 
with $N$ mean photons per mode, that is, $\rho^n_{AA'} = \rho^{\otimes n}_{AA'}$
with 
\begin{equation}
\rho_{AA'} = |\zeta_N\rangle_{AA'}\langle\zeta_N|
\end{equation}
and 
\begin{equation}
|\zeta_N\rangle_{AA'} = \frac{1}{\sqrt{N+1}} \sum_{\ell=0}^\infty \left( \frac{N}{N+1} \right)^{\ell/2} |\ell\rangle_{A}|\ell\rangle_{A'} \, ,
\end{equation}
where $|\ell\rangle$ denotes the photon-number state with $\ell$ photons. 
Alice keeps the modes labeled with `$A$' and sends through $n$ uses of a lossy bosonic channel those labeled with `$A'$',
see Fig.\ \ref{fig:dreverse}.
At the end of this first phase of the communication protocol, Alice, Bob and Eve share the $3n$-mode 
state $\rho^n_{ABE} = \rho^{\otimes n}_{ABE}$, where $\rho_{ABE}$ is a $3$-mode Gaussian state
with zero mean and covariance matrix $V_{ABE}$ (whose explicit form is given in the Appendix).

In the second phase of the communication protocol, Bob makes a collective measurement
on his share of $n$ bosonic modes, described by the state $\rho^n_B = \rho_B^{\otimes n}$,
where $\rho_B$ is a Gaussian state with zero mean and variance $V_B$ (see Appendix for details).
Indeed, Bob applies a measurement $\Gamma(s)$ chosen from a set of measurements
parameterized by the label $s = 1, \dots, K$.
The value of $s$ is determined by the secret key he shares with Alice.
That is, while the list of possible $K$ measurement is public and hence known to
Eve, the specific choice of $\Gamma(s)$ is known only by Alice and Bob.

Bob's measurement is defined as follows. 
First, Alice and Bob publicly agree on a set of $MK$ $n$-mode coherent states
\begin{equation}
|\beta^n(x,s)\rangle = \bigotimes_{j=1}^n |\beta_j(x,s)\rangle \, ,
\end{equation}
for $x = 1, \dots, M$ and $s = 1, \dots, K$. 
These coherent states are defined by sampling the
amplitudes $\beta_j(x,s)$ i.i.d.\ from a circularly symmetric Gaussian distribution with zero mean and
variance $\eta N$. 
For any given $s$, we consider the sliced operator
\begin{equation}
\Sigma(s) = \sum_{x=1}^{M} \mathbb{P}^n_B \, |\beta^n(x,s)\rangle \langle \beta^n(x,s)| \, \mathbb{P}^n_B \, ,
\end{equation}
where $\mathbb{P}^n_B$ is the projector on the strongly $\delta$-typical subspace defined by $\rho_B^{\otimes n}$ (see, e.g.\ \cite{Wilde}).
Applying the operator Chernoff bound (see Appendix for details) we obtain that the bounds
\begin{equation}
(1-\epsilon) M 2^{-n g(\eta N)} \mathbb{P}^n_B \leq \Sigma(s) \leq (1+\epsilon) M 2^{-n g(\eta N)} \mathbb{P}^n_B
\end{equation}
hold true with arbitrarily high probability provided $M \gg 2^{n g(\eta N)}$.
It follows that for any given $s$ the operators
\begin{equation}\label{gamma}
\Gamma_{x}(s) = \frac{ \mathbb{P}^n_B \, |\beta^n(x,s)\rangle \langle \beta^n(x,s)| \, \mathbb{P}^n_B }{(1+\epsilon) M 2^{-n g(\eta N)} } 
\end{equation}
define a subnormalized POVM in Bob's typical subspace, which can be completed by
introducing the operator $\Gamma_{0}(s) = \mathbb{P}^n_B - \sum_{x} \Gamma_{x}(s)$.
In this way we have defined Bob's measurement $\Gamma(s)$ for all values of $s$.
After performing the measurement, Bob declares an error if he obtains the 
measurement output corresponding to $\Gamma_{0}(s)$. This event, however, happens
with a negligible probability (see Appendix for details).

In the third phase of the protocol, Alice makes a measurement on her share of bosonic modes.
For a given value of $s$ (which is known to Alice and Bob) and $x$, we consider Alice's 
conditional state $\rho^n_A(x,s)$.
As a matter of fact, Bob's measurement induces a virtual backward communication channel
from Bob to Alice.
As a result, for given $s$, Alice obtains an ensemble of states $\{ \rho^n_A(x,s) , p(x,s) \}_{x=1,\dots,M}$,
where $p(x,s) = \mathrm{Tr}(\Gamma_{x}(s) \rho^n_B(s))$.
The maximum amount of classical information (per mode) about $x$ that Alice can extract from this ensemble of states
is given, in the asymptotic setting, by the associated Holevo information \cite{Holevo} \cite{NOTA_superb}:
\begin{equation}
\chi_\mathrm{rr} = \frac{1}{n} \left[ S(\rho^n_A) - \sum_{x} p(x,s) S(\rho^n_A(x,s)) \right] \, ,
\end{equation}
where $S(\rho) = - \mathrm{Tr}\left( \rho \log{\rho}\right)$ denotes the von Neumann entropy.
From the explicit expressions for $p(x,s)$, $\rho^n_A(x,s)$ and $\rho^n_A$ (given in the Appendix)
we obtain
\begin{equation}
\chi_\mathrm{rr} = g(N) - g[(1-\eta)N']
\end{equation}
where $N' = N/(1+\eta N)$.
$\chi_\mathrm{rr}$ also quantifies the rate (in bits per mode) of shared randomness
that can be established, with the assistance of public communication, by Alice and Bob \cite{errcorr}.

%--

Finally, to show the security of the communication protocol, we need to bound Eve's 
accessible information about $x$.
Bob's measurement also induces a virtual quantum channel to Eve.
For any given $s$, the ensemble of states obtained by Eve is $\{ \rho^n_E(x,s) , p(x,s) \}_{x=1,\dots,M}$,
where $\rho^n_E(x,s)$ is Eve's state conditioned on Bob's measurement result $x$.
Given the explicit form of $\rho^n_E(x,s)$ we show (see Sec.\ \ref{sec:proofs} and the Appendix) that Eve's accessible
information about $x$ is negligibly small for $K$ such that
\begin{align}
k_\mathrm{rr} & := \lim_n \frac{\log{K}}{n } \\
& = 2 g[(1-\eta)N] - g[(1-\eta)N'] - g[(1-\eta)N''] \, ,
\end{align}
with $N'' = N(1+2\eta N)/(1+\eta N)$.
In conclusion, applying the bootstrapping routine, we obtain a net rate of locked-key generation of (in bits per mode) 
\begin{equation}
r_\mathrm{rr} = \chi_\mathrm{rr} - k_\mathrm{rr}
= g(N) - 2 g[(1-\eta)N] + g[(1-\eta)N''] \, ,
\end{equation}
which in the limit of $N \to \infty$ reads
\begin{equation}
r_\mathrm{rr} = 1 + \log{\left( \frac{1}{1-\eta} \right) } \, .
\end{equation}

Similar results are obtained if the channel from Alice to Bob is
lossy and noisy. In this case the reverse reconciliation protocol 
achieves an asymptotic locked-key rate of
\begin{equation}
r_\mathrm{rr} = 1 + \log{\left( \frac{1}{1-\eta} \right) } - g(N_T) \, ,
\end{equation}
where $N_T$ is the mean number of thermal photons per mode in the channel.

\section{Security proofs}\label{sec:proofs}

We discuss in details the case of the lossy channel. 
The proof for the lossy and noisy channel can be obtained in a similar way.

The starting point of the proof are some mathematical tools presented in \cite{NJP}.
There we assumed that Eve's states $\rho^n_E(x,s)$ belongs to a finite-dimensional
space of dimension $d^n$.
Given the bipartite state 
\begin{equation}
\rho^n_{XE} = \frac{1}{M} \sum_{x=1}^M |x\rangle\langle x| \otimes \frac{1}{K}\sum_{s=1}^K \rho^n_E(x,s) \, ,
\end{equation}
the following bound hold for the associated accessible information (see~\cite{NJP}):
\begin{equation}\label{Iaccn1}
I_\mathrm{acc}%(\rho^n_{XE}) 
\leq \log{M} - \frac{d^n}{M} \, \min_{|\phi\rangle} \left\{  H[Q(\phi)] - \eta\left[ \sum_{x=1}^M Q_x(\phi)\right] \right\} \, ,
\end{equation}
where 
\begin{equation}
Q_x(\phi) = \frac{1}{K} \sum_{s=1}^K \langle \phi | \rho^n_E(x,s) | \phi \rangle \, ,
\end{equation}
$H[Q(\phi)] = \sum_{x=1}^M \eta( Q_x(\phi) )$,
with $\eta( \, \cdot \, ) = - ( \, \cdot \, ) \log{( \, \cdot \, )}$.
The minimum is over all vectors $\phi$ in Eve's $d^n$-dimensional Hilbert space.

As shown in \cite{NJP}, if the ensemble of states from which the codewords 
are sampled is such that for any unit vector $|\phi\rangle$, 
\begin{equation}\label{1moment}
\mu := \mathbb{E}_s[ \langle \phi | \rho^n_E(x,s) | \phi \rangle ] = \frac{1}{d^n} 
\end{equation}
($\mathbb{E}_s$ denotes the expectation value over $s$), 
and 
\begin{align}
\Sigma & := \mathbb{E}_s[ \langle \phi | \rho^n_E(x,s) | \phi \rangle^2] \nonumber \\
& = \mathbb{E}_s[ \langle \phi,\phi | \rho^n_E(x,s) \otimes \rho^n_E(x,s) | \phi,\phi \rangle ] \label{2moment} \, ,
\end{align}
(here $|\phi,\phi\rangle \equiv |\phi\rangle \otimes |\phi\rangle$) 
then the right hand side of (\ref{Iaccn1}) is smaller than $\epsilon \log{M}$
provided that 
\begin{equation}\label{Kqdl}
K > \max\left\{ 2 \gamma^n  \left( \frac{1}{\epsilon^2} \ln{M} + \frac{2}{\epsilon^3} \ln{\frac{5}{\epsilon}}\right) , 
\frac{d^n}{M} \, \frac{4 \ln{2} \ln{d^n}}{\epsilon^2 }
\right\} \, ,
\end{equation}
with
\begin{equation}
\gamma^n = \frac{\Sigma}{\mu^2} \, .
\end{equation}

In our setting $n$ counts the number of modes employed in one quantum data locking routine. 
Putting $M = 2^{n \chi}$ and $\epsilon = e^{-n^c}$ with $c \in (0,1)$, condition
(\ref{Kqdl}) yields an asymptotic rate of secret-key consumption (in bits per mode)
\begin{equation}
k = \lim_{n\to\infty} \frac{1}{n} \log{K} =  \max\left\{ \log{\gamma} , \log{d}-\chi \right\} \, .
\end{equation}

%---

In our continuous-variable setting, Eve's space is infinite-dimensional.
Therefore, to apply the result of \cite{NJP} we need to map Eve's space into
a finite dimensional one.
In both the direct and reverse reconciliation protocol, the expectation value over $s$ 
of the state of Eve has the form (see details in the Appendix)
\begin{equation}
\rho^n_E = \mathbb{E}_s [ \rho_E^n(x,s) ] = \rho_E^{\otimes n} \, ,
\end{equation}
that is, the average state is a direct product.
In particular, $\rho_E$ is a Gaussian state with zero mean, variance $V_E$, 
and mean photon number $(1-\eta) N$.
We can hence consider the $\delta$-typical subspace projector $\mathbb{P}^n_\rho$ 
associated with $\rho_E^{\otimes n}$.
We use this projector to define an auxiliary bipartite state of the form
\begin{equation}
\sigma^n_{XE} = \frac{1}{M} \sum_{x=1}^M |x\rangle\langle x| \otimes \frac{1}{K}\sum_{s=1}^K \sigma^n_E(x,s) \, ,
\end{equation}
where
\begin{equation}
\sigma^n_E(x,s) = \mathbb{P}^n_\rho \, \rho^n_E(x,s) \, \mathbb{P}^n_\rho
\end{equation}
is obtained by slicing with the $\delta$-typical subspace projector.
From the properties of the typical projector we have
\begin{equation}
\| \sigma^n_{XE} - \rho^n_{XE} \|_1 \leq \delta \, .
\end{equation}
Since the two states are $\delta$-close in trace-norm, the security of the
state $\rho^n_{XE}$ follows, up to a probability $\delta$, from that
of $\sigma^n_{XE}$.
In such a way we have reduced the problem to a finite dimensional one,
where the dimension is that of the $\delta$-typical subspace, i.e.,
\begin{equation}
d^n := \mathrm{Tr}\left( \mathbb{P}^n_\rho \right) \in [ 2^{n [S(\rho_E) - c \delta]} , 2^{n [S(\rho_E) + c \delta]}]
\end{equation}
(for some constant $c$).

We use a notion of typical subspace that is a slightly different 
from the one usually considered (see for instance \cite{Wilde}).
Given a hermitian operator $\xi$ we consider its spectral decomposition
\begin{equation}
\xi = \sum_\ell p_\ell \, P_\ell \, ,
\end{equation}
where the sum is over the eigenvalues $p_\ell$ and the corresponding eigenprojectors $P_\ell$,
in such a way that $p_\ell \neq p_{\ell'}$ for $\ell \neq \ell'$ (that is, $\mathrm{Tr} \left( P_\ell \right)$
equals the degeneracy of $p_\ell$).
We look at each projector $P_\ell$ as an event whose probability is $\pi_\ell = p_\ell \, \mathrm{Tr} \left( P_\ell \right)$.
Given $\xi^{\otimes n}$, we then define the $\delta$-typical projector $\mathbb{P}^n_\xi$ as
(we omit the subscript $\delta$ to simplify the notation)
\begin{equation}
\mathbb{P}^n_\xi = \sum_{p_{\ell_1}p_{\ell_2}\cdots p_{\ell_n} \in T_\delta^n} P_{\ell_1} \otimes P_{\ell_2} \otimes \cdots \otimes P_{\ell_n} 
\end{equation}
where the sum is over the sequences $p_{\ell_1}p_{\ell_2}\cdots p_{\ell_n}$ which are 
$\delta$-typical with respect to the probability distribution $\pi_\ell$. 
Notice that this construction of the typical projector coincides with the
usual one when all the eigenvalues of $\xi$ are non degenerate.

First we compute (\ref{1moment}):
\begin{align}
\mu & = \mathbb{E}_s[ \langle \phi | \, \sigma^n_E(x,s) \, | \phi \rangle ] \\
& = \mathbb{E}_s[ \langle \phi | \, \mathbb{P}^n_\rho \, \rho^n_E(x,s) \, \mathbb{P}^n_\rho \, | \phi \rangle ] \\
& = \langle \phi | \, \mathbb{P}^n_\rho \, \rho_E^{\otimes n} \, \mathbb{P}^n_\rho \, | \phi \rangle \, .
\end{align}
Then, from the equipartition properties of the $\delta$-typical subspace we have
(for some constant $c$)
\begin{equation}
2^{-n [ S(\rho_E) + c\delta ]} \leq \mu \leq 2^{-n [ S(\rho_E) - c\delta]} \, .
\end{equation}

To compute (\ref{2moment}) we need to introduce another typical subspace projector.
We consider the state $( \rho_E \otimes \rho_E)^{\otimes n}$ and its associated
$(2\delta)$-typical subspace projector, denoted as $\mathbb{P}^n_{\rho \otimes \rho}$.
Notice that $[ \mathbb{P}_\rho^n \otimes \mathbb{P}_\rho^n , \mathbb{P}^n_{\rho \otimes \rho} ] = 0$, and that
$\mathbb{P}_\rho^n \otimes \mathbb{P}_\rho^n \leq \mathbb{P}^n_{\rho \otimes \rho}$.

We also consider the state 
\begin{equation}
\rho_{2E}^n := \mathbb{E}_s[ \rho^n_E(x,s) \otimes \rho^n_E(x,s) ] = \rho_{2E}^{\otimes n} \, .
\end{equation}
By explicit computation (see Appendix) we can show that, in
both the direct and reverse reconciliation protocols, $\rho_{2E}$ is a Gaussian state
with zero mean and covariance matrix $V_{2E}$.
Moreover, $\rho_{2E}$ commutes with $\rho_E \otimes \rho_E$
since they are both diagonal in the photon-number basis (see Appendix). 
It follows that $\rho_{2E}$ also commutes with $\mathbb{P}^n_{\rho \otimes \rho}$.
We also have that, given that $\rho_E$ has mean photon number $(1-\eta)N$, then both 
$\rho_{2E}$ and $\rho_E \otimes \rho_E$ have $2(1-\eta)N$ mean photons.

We can now compute (\ref{2moment}):
\begin{align}
\Sigma & = \mathbb{E}_s[ \langle \phi, \phi | \, \sigma^n_E(x,s) \otimes \sigma^n_E(x,s) \, | \phi, \phi \rangle ] \\
& = \mathbb{E}_s[ 
\langle \phi, \phi | \, {\mathbb{P}_\rho^n}^{\otimes 2} \, \rho^n_E(x,s) \otimes \rho^n_E(x,s) \, {\mathbb{P}_\rho^n}^{\otimes 2} \, | \phi, \phi \rangle
] \\
& = \langle \phi, \phi | \, {\mathbb{P}_\rho^n}^{\otimes 2} \, \rho_{2E}^{\otimes n} \, {\mathbb{P}_\rho^n}^{\otimes 2} \, | \phi, \phi \rangle \, .
\end{align}
Since ${\mathbb{P}_\rho^n}^{\otimes 2}$ commutes with $\mathbb{P}_{\rho\otimes\rho}^n$ and
${\mathbb{P}_\rho^n}^{\otimes 2} \leq \mathbb{P}_{\rho\otimes\rho}^n$, we have
\begin{align}
\Sigma & \leq \langle \phi, \phi | \,
\mathbb{P}_{\rho\otimes\rho}^n \, \rho_{2E}^{\otimes n} \, \mathbb{P}_{\rho\otimes\rho}^n
\, | \phi, \phi \rangle \, .
\end{align}

To conclude, let us consider the sliced operator
$\mathbb{P}_{\rho\otimes\rho}^n \, \rho_{2E}^{\otimes n} \, \mathbb{P}_{\rho\otimes\rho}^n$.
Since $[ \mathbb{P}_{\rho\otimes\rho}^n , \rho_{2E}^{\otimes n} ] = 0$ we can apply
a classical argument concerning typical type classes (see, e.g., \cite{CT}).
Let us denote as $q_\ell$ the eigenvalues of $\rho_{2E}$. 
We notice that the eigenvectors of $\mathbb{P}_{\rho\otimes\rho}^n \, \rho_{2E}^{\otimes n} \, \mathbb{P}_{\rho\otimes\rho}^n$
are those of $\rho_{2E}^{\otimes n}$ which are in the range of $\mathbb{P}_{\rho\otimes\rho}^n$
(that is, they are $\delta$-typical for $(\rho_E \otimes \rho_E)^{\otimes n}$).
Consider then an eigenvector whose $\delta$-typical type is $\tilde\pi$, the corresponding
eigenvalue of $\mathbb{P}_{\rho\otimes\rho}^n \, \rho_{2E}^{\otimes n} \, \mathbb{P}_{\rho\otimes\rho}^n$ is 
\begin{equation}
w = \prod_\ell q_\ell^{n \tilde\pi_\ell} = 2^{n \sum_\ell \tilde\pi_\ell \log{q_\ell}} \, .
\end{equation}
Being $\rho_{2E}$ a zero-mean, thermal-like, Gaussian state, $q_\ell = Z^{-1} 2^{-\beta \ell}$, where $\ell$ 
is the photon number. This yields
\begin{align}
w & = 2^{- n \left(  \beta \langle \ell \rangle_{\tilde\pi} + \log{Z} \right)} \\
& = 2^{- n \left(  \beta 2(1-\eta)N  + \log{Z} + \beta \Delta \langle \ell \rangle  \right)} \\
& = 2^{- n \left(  S(\rho_{2E}) + \beta \Delta \langle \ell \rangle  \right)} \, .
\end{align}
Here $\langle \ell \rangle_{\tilde\pi} = \sum_\ell \tilde\pi_\ell \ell$ is the mean photon
number given by the $\delta$-typical distribution $\tilde\pi$.
Since $\tilde\pi$ is $\delta$-typical for $(\rho_E \otimes \rho_E)^{\otimes n}$,
we expect $\langle \ell \rangle_{\tilde\pi} = 2(1-\eta)N$,
$\Delta \langle \ell \rangle = \langle \ell \rangle_{\tilde\pi} - 2(1-\eta)N$
being the fluctuation about the expectation value.
Finally we have used 
$S(\rho_{2E}) = \beta 2(1-\eta)N + \log{Z}$.
In the Appendix we show that, for a $\delta$-typical type $\tilde\pi$, 
\begin{equation}
| \beta \Delta \langle \ell \rangle | \leq 2 c \delta [ (1-\eta)N + 1]
\end{equation} 
(for some constant $c$), from which we obtain
\begin{equation}
w \leq 2^{- n \left(  S(\rho_{2E}) + 2 c \delta [(1-\eta)N +1] \right)} \, ,
\end{equation}
and hence
\begin{equation}
\Sigma \leq 2^{- n \left(  S(\rho_{2E}) + 2 c \delta [(1-\eta)N +1] \right)} \, .
\end{equation}

From these results, in the limits that $n \to \infty$ and $\delta \to 0$,
we obtain the following bound on the key consumption rate:
\begin{equation}
k = \max\left\{   
2 S(\rho_E) - S(\rho_{2E}) ,
S(\rho_E) - \chi
\right\} \, .
\end{equation}

For the direct reconciliation protocol we have (see derivation in Appendix):
$\chi = g(\eta N)$,
$S(\rho_E) = g((1-\eta) N)$, and
$S(\rho_{2E}) = g(2(1-\eta) N)$. 
For any given $\eta > 0$ and $N$ large enough we then obtain
\begin{equation}
k = 2 S(\rho_E) - S(\rho_{2E}) = 2 g[(1-\eta)N] - g[2(1-\eta)N] \, .
\end{equation}

For the reverse reconciliation protocol we have (see Appendix)
$\chi = g(N) - g((1-\eta) N')$, with $N' = N/(1+\eta N)$,
$S(\rho_E) = g((1-\eta) N)$, and 
$S(\rho_{2E}) = g((1-\eta) N') + g((1-\eta) N'')$, with $N'' = N ( 1 + 2 \eta N )/( 1 + \eta N)$.
For any given $\eta > 0$ and $N$ large enough we then obtain
\begin{align}
k & = 2 S(\rho_E) - S(\rho_{2E}) \\
& = 2 g[(1-\eta)N] - g[(1-\eta)N'] - g[(1-\eta)N''] \, .
\end{align}

%--

\section{Conclusion}

Quantum cryptography promises unconditionally secure communication through insecure communication channels.
However, fundamental properties of quantum entanglement bound the ultimate
secret-key generation rates that can be achieved through a communication channel \cite{TGW}.
For the relevant case of a lossy communication line, as e.g. free-space of fiber optics
communication, the bound of \cite{TGW} implies that the secret-key generation rate must
decrease at least exponentially with increasing communication distance.

Here we have analyzed the rate-distance tradeoff under the realistic assumption that
one can store quantum information reliably only for a finite time. Clearly, any quantum
memory device can store quantum information only for a time of the order of its coherence
time. We have shown that for any given finite, yet arbitrarily long, storage time, the 
quantum data locking effect can be applied to generate key at a constant rate over 
arbitrarily long distances through an optical channel with linear loss. 
Moreover, we have shown that this result holds also in the presence of moderate noise or experimental
imperfections modeled as a thermal background.

It remains an open problem to show that these high rates of key generation can be achieved in practice.
One major problem is to find a decoding measurement that can be experimentally 
realized with current technologies and still allows us to achieve a constant key 
rate over long communication distances.
If this question will find a positive answer, our results could pave the way
to a new family of QKD protocols that yield a constant key 
rate that does {\it not} decay with increasing communication distance.
This would also imply that long distance quantum communication can be in principle
realized without employing quantum repeaters.

\acknowledgments

We are grateful to Bhaskar Roy, Mark M. Wilde, Saikat Guha, and Hari Krovi for valuable discussions and suggestions. 
This research was supported by the DARPA Quiness Program through U.S. Army Research Office Grant No. W31P4Q-12-1-0019.
CL was supported by the SUTD--MIT Graduate Fellows Program.

\appendix

\section{The direct reconciliation protocol}

In the direct reconciliation protocol, the $n$-mode codewords obtained by Eve read
\begin{equation}
\rho^n_E(x,s) = |\sqrt{1-\eta} \, \alpha^n(x,s) \rangle \langle \sqrt{1-\eta} \, \alpha^n(x,s) | \, ,
\end{equation}
where $|\sqrt{1-\eta} \, \alpha^n(x,s) \rangle = \otimes_{j=1}^n |\sqrt{1-\eta}\, \alpha_j(x,s) \rangle$ is a $n$-mode coherent
state, where the amplitudes $\alpha_j(x,s)$'s are sampled i.i.d.\ from a circularly symmetric Gaussian distribution 
$G_{(0,N)} = \frac{1}{2\pi N} \, e^{-|\alpha|^2/N}$ 
with zero mean and variance $N$. 
Therefore the expectation value over $s$ of $\rho^n_E(x,s)$ reads 
\begin{align}
\mathbb{E}_s[\rho^n_E(x,s)] & = \left( 
\int d\mu
|\sqrt{1-\eta} \, \alpha \rangle \langle \sqrt{1-\eta} \, \alpha | 
\right)^{\otimes n} \\
& = \rho_{E}^{\otimes n} \, ,
\end{align}
where $d\mu = d^2\alpha \, G_{(0,N)}(\alpha)$, and 
$\rho_{E}$ is a single-mode thermal state with mean photon number $(1-\eta)N$.
The spectral decomposition of $\rho_{E}$ is
\begin{equation}\label{spectre1}
\rho_{E} = \frac{1}{(1-\eta)N+1} \sum_{\ell=0}^\infty \left( \frac{(1-\eta)N}{(1-\eta)N+1} \right)^\ell |\ell\rangle \langle \ell| \, ,
\end{equation}
where $|\ell\rangle$ is the $\ell$-photon state.
The von Neumann entropy of $\rho_E$ is
\begin{equation}
S(\rho_E) = g((1-\eta)N) \, .
\end{equation}
Therefore, denoting as $\mathbb{P}_\rho^n$ the $\delta$-typical projector associated with $\rho_E^n$
we have (for some constant $c$) (see e.g.\ \cite{Wilde}) 
\begin{equation}
2^{n [g((1-\eta)N) - c\delta]} \leq \mathrm{Tr}(\mathbb{P}_\rho^n) \leq 2^{n [g((1-\eta)N) + c\delta]}
\end{equation}
and 
\begin{equation}
2^{- n [g((1-\eta)N) + c\delta]} \mathbb{P}_\rho^n \leq \mathbb{P}_\rho^n \, \rho_E^n \, \mathbb{P}_\rho^n \leq 2^{-n [g((1-\eta)N) - c\delta]} \mathbb{P}_\rho^n \, .
\end{equation}

Consider the operator $\rho_E^{\otimes 2}$.
This is a two-mode thermal state with $2(1-\eta) N$ mean photons.
Its spectral decomposition can be obtained from (\ref{spectre1}):
\begin{equation}\label{spectre2}
\rho_E^{\otimes 2} = \left( \frac{1}{(1-\eta)N+1} \right)^2 \sum_{\ell=0}^\infty \left( \frac{(1-\eta)N}{(1-\eta)N+1} \right)^\ell P_\ell \, ,
\end{equation}
where $P_\ell$ denotes the projector on the subspace with $\ell$ photons.
The $\ell$-photon subspace is generated by the $\ell +1$ two-mode vectors 
$\{ |0\rangle|\ell \rangle, |1\rangle|\ell-1 \rangle, \dots |\ell \rangle|0 \rangle\}$,
therefore $\mathrm{Tr}\left( P_\ell \right) = \ell+1$.

Let us now consider the expectation value over $s$ of the operator $\rho^n_E(x,s) \otimes \rho^n_E(x,s)$:
\begin{align}
& \mathbb{E}_s[ \rho^n_E(x,s) \otimes \rho^n_E(x,s) ] = \nonumber \\
& \left( \int d\mu
|\sqrt{1-\eta} \alpha \rangle \langle \sqrt{1-\eta} \alpha| \otimes |\sqrt{1-\eta} \alpha \rangle \langle \sqrt{1-\eta} \alpha| 
\right)^{\otimes n} \\
& = \rho_{2E}^{\otimes n} \, .
\end{align}
The state $\rho_{2E}$ is a Gaussian state with zero mean and $2(1-\eta)N$ mean photons.
Its spectral decomposition is:
\begin{equation}
\rho_{2E} = 
\frac{1}{2(1-\eta)N+1} 
\sum_{\ell=0}^\infty \left( \frac{2(1-\eta)N}{2(1-\eta)N+1} \right)^\ell |\psi^+_\ell\rangle \langle \psi^+_\ell| \, ,
\end{equation}
where
\begin{equation}
|\psi^+_\ell\rangle = 2^{-\ell/2} \sum_{i=0}^\ell \sqrt{ { \ell \choose i}} \, | i \rangle | \ell-i \rangle \, .
\end{equation}
From this we compute the von Neumann entropy of $\rho_{2E}$:
\begin{equation}
S(\rho_{2E}) = g(2(1-\eta)N) \, .
\end{equation}
Finally, since $|\psi_\ell\rangle$ is a $\ell$-photon state, we obtain that 
$\rho_{2E}$ commutes with $\rho_E \otimes \rho_E$, which also implies that 
$[ \rho_{2E}^{\otimes n} , \mathbb{P}_{\rho \otimes \rho}^n ] = 0$.

\section{The reverse reconciliation protocol}

\subsection{Bob's measurement}\label{ss1}

We recall the statement of the operator Chernoff bound \cite{Chernoff}.
Let $\{ \xi_t \}_{t=1,\dots,T}$ be a collection of i.i.d.\ operator-valued random variables,
where each $\xi_t$ is a positive hermitian operator in a Hilbert space of dimension $D$,
satisfying $\xi_t \leq \mathbb{I}$ and with mean value $\mathbb{E}[\xi_t] = \mu \geq a \mathbb{I}$
for some $a \in (0,1)$.
Then for any $\epsilon > 0$ (and provided that $(1+\epsilon) \mu < 1$) we have 
\begin{equation}\label{CB1}
\mathrm{Pr}\left\{ \frac{1}{T} \sum_{t=1}^T \xi_t \geq (1+\epsilon) \mu \right\} \leq D \exp{\left( - \frac{T \epsilon^2 a}{4 \ln{2}} \right)} \, ,
\end{equation}
and
\begin{equation}\label{CB2}
\mathrm{Pr}\left\{ \frac{1}{T} \sum_{t=1}^T \xi_t \leq (1-\epsilon) \mu \right\} \leq D \exp{\left( - \frac{T \epsilon^2 a}{4 \ln{2}} \right)} \, ,
\end{equation}

To define Bob's POVM we apply this bound to the operators
\begin{align}
\xi(x,s) = \mathbb{P}_B^n \, |\beta^n(x,s) \rangle \langle \beta^n(x,s)| \, \mathbb{P}_B^n \, ,
\end{align}
where $\mathbb{P}_B^n$ is the projector on Bob's typical subspace.
For any given $s$, we have a collection $\{ \xi(x,s) \}_{x=1,\dots,M}$ of $M$ i.i.d.\ operator-valued random variables,
with $\xi(x,s) \leq \mathbb{P}^n_B$, and 
$\mathbb{E}[\xi(x,s)] \geq 2^{-n [ g(\eta N) + c \delta]} \mathbb{P}^n_B$.
Hence by restricting to Bob's typical subspace, we meet the conditions for applying the 
operator Chernoff bound with $a = 2^{-n [ g(\eta N) + c \delta]}$.
It follows from (\ref{CB1}) that for any $s$, the operator
\begin{equation}
\Sigma(s) = \sum_{x=1}^{M} \mathbb{P}^n_B \, |\beta^n(x,s)\rangle \langle \beta^n(x,s)| \, \mathbb{P}^n_B 
\end{equation}
satisfies $\Sigma(s) \leq M (1+\epsilon) 2^{-n [ g(\eta N) + c \delta]} \mathbb{P}^n_B$
with arbitrary high probability if $M \gg 2^{-n [ g(\eta N) + c \delta]}$.
This in turn implies that the operators 
\begin{equation}
\Gamma_{x}(s) = \frac{ \mathbb{P}^n_B \, |\beta^n(x,s)\rangle \langle \beta^n(x,s)| \, \mathbb{P}^n_B }{(1+\epsilon) M 2^{-n [ g(\eta N) + c \delta]} } 
\end{equation}
define a subnormalized POVM, that is, $\sum_x \Gamma_x(s) \leq \mathbb{I}$
(here the identity is intended as the identity operator in the typical subspace).

To complete the subnormalized POVM we introduce the operator
\begin{equation}
\Gamma_{0}(s) = \mathbb{I} - \sum_x \Gamma_{0}(s) \, .
\end{equation}
However, that the probability associated to
the POVM element $\Gamma_{0}(s)$ is negligibly small.
Applying (\ref{CB2}) we obtain that 
\begin{equation}
\sum_x \Gamma_{x}(s) \geq \frac{1-\epsilon}{1+\epsilon} \simeq 1-\epsilon^2 \, ,
\end{equation}
from which it follows $\Gamma_{0}(s) \lesssim \epsilon^2$.

\subsection{Alice's and Eve's conditional states}\label{ss2}

For the reverse reconciliation protocol it is easier to work in the
Wigner function representation.

In the first phase of the reverse reconciliation protocol
the tripartite state $\rho_{ABE}^{n} = \rho_{ABE}^{\otimes n}$ is broadcast
by Alice through the quantum channel. 
$\rho_{ABE}^{\otimes n}$ is the tensor product of $n$ three-mode zero-mean Guassian states
(for a review on Gaussian states see, e.g., \cite{biblio}).
The Wigner function of $\rho_{ABE}$ reads 
\begin{equation}
W(\mathbf{R}_{ABE}) = \mathcal{N} \exp{\left( - \frac{1}{2} \, \mathbf{R}_{ABE} V_{ABE}^{-1} \mathbf{R}^\mathsf{T}_{ABE}\right)} \, ,
\end{equation}
where $\mathbf{R}_{ABE} = (q_A,p_A , q_B,p_B , q_E,p_E)$ is the three-mode quadrature vector.
The covariance matrix can be easily computed and reads:
\begin{widetext}
\begin{align}\label{CM-lossy}
V_{ABE} = \frac{1}{2} \left(
\begin{array}{cccccc}
C              & 0               & S\sqrt{\eta}             & 0                        & S\sqrt{1-\eta}           & 0 \\
0              & C               & 0                        & -S\sqrt{\eta}            & 0                        & -S\sqrt{1-\eta} \\
S\sqrt{\eta}   & 0               & C\eta+(1-\eta)           & 0                        & (C-1)\sqrt{\eta(1-\eta)} & 0 \\
0              & -S\sqrt{\eta}   & 0                        & C\eta+(1-\eta)           & 0                        & (C-1)\sqrt{\eta(1-\eta)} \\
S\sqrt{1-\eta} & 0               & (C-1)\sqrt{\eta(1-\eta)} & 0                        & C(1-\eta)+\eta           & 0 \\
0              & -S\sqrt{1-\eta} & 0                        & (C-1)\sqrt{\eta(1-\eta)} & 0                        & C(1-\eta)+\eta
\end{array}
\right)
\end{align}
where $C = 2N+1$ and $S = 2 \sqrt{N(N+1)}$.
From $V_{ABE}$ we obtain the covariance matrix of the joint state of Alice and Bob,
\begin{eqnarray}
V_{AB} = \frac{1}{2} \left(
\begin{array}{cccc}
C              & 0               & S\sqrt{\eta}             & 0                        \\
0              & C               & 0                        & -S\sqrt{\eta}            \\
S\sqrt{\eta}   & 0               & C\eta+(1-\eta)           & 0                        \\
0              & -S\sqrt{\eta}   & 0                        & C\eta+(1-\eta)           
\end{array}
\right)
\end{eqnarray}
and that of Eve and Bob,
\begin{eqnarray}\label{BE-CM}
V_{BE} = \frac{1}{2} \left(
\begin{array}{cccc}
C\eta+(1-\eta)           & 0                        & (C-1)\sqrt{\eta(1-\eta)} & 0 \\
0                        & C\eta+(1-\eta)           & 0                        & (C-1)\sqrt{\eta(1-\eta)} \\
(C-1)\sqrt{\eta(1-\eta)} & 0                        & C(1-\eta)+\eta           & 0 \\
0                        & (C-1)\sqrt{\eta(1-\eta)} & 0                        & C(1-\eta)+\eta
\end{array}
\right) \, .
\end{eqnarray}
\end{widetext}

In the second phase of the protocol Bob makes a measurement described by the
POVM elements $\Gamma_x(s)$ (\ref{gamma}). To simplify the notation we drop the normalization
factor and write
\begin{equation}
\Gamma_x(s) \simeq \mathbb{P}^n_B \, |\beta^n(x,s) \rangle \langle \beta^n(x,s)| \, \mathbb{P}^n_B \, .
\end{equation}

We compute Alice's (not-normalized) conditional state:
\begin{align}
\rho^n_A(x,s) & = \mathrm{Tr}_{B} [ \mathbb{I}_A^n \otimes \Lambda_x^{(s)} \rho^{\otimes n}_{AB} ] \\
& = \mathrm{Tr}_{B} \left[ \mathbb{I}_A^n \otimes \mathbb{P}^n_B \, |\beta^n(x,s) \rangle \langle \beta^n(x,s)| \, \mathbb{P}^n_B \, \rho^{\otimes n}_{AB} \right] \, .
\end{align}
We apply the property of strong typicality,
$\| \mathbb{P}^n_B \, |\beta^n(x,s) \rangle \langle \beta^n(x,s)| \, \mathbb{P}^n_B - |\beta^n(x,s) \rangle \langle \beta^n(x,s)| \|_1 \leq \delta$,
to obtain, up to an error smaller than $\delta$ in trace distance,
\begin{align}
\rho^n_A(x,s) & \simeq \mathrm{Tr}_{B} \left[ \mathbb{I}_A^n \otimes |\beta^n(x,s) \rangle \langle \beta^n(x,s)| \, \rho^{\otimes n}_{AB} \right] \\
& = \bigotimes_{j=1}^n \mathrm{Tr}_{B} \left[ \mathbb{I}_A \otimes |\beta_j(x,s) \rangle \langle \beta_j(x,s)| \, \rho_{AB} \right] \\
& = \bigotimes_{j=1}^n \rho_{A_j}(x,s) \, .
\end{align}
Then the probability of the outcome `$x$' can be obtained as $p(x,s) = \mathrm{Tr}\left[ \rho^n_A(x,s) \right]$.

In the Wigner function representation, the equation 
$\rho_{A_j}(x,s) = \mathrm{Tr}_{B} \left[ \mathbb{I}_A  \otimes |\beta_j(x,s) \rangle \langle \beta_j(x,s)| \, \rho_{AB} \right]$
reads
\begin{equation}
W_{A_j(x,s)}(\mathbf{R}_A) = (2\pi)^n \int d^{2n} \mathbf{R}_B W_{\beta_j(x,s)}(\mathbf{R}_B) W_{AB}(\mathbf{R}_{AB}) \, ,
\end{equation}
where $W_{AB}(\mathbf{R}_{AB})$ is the Wigner function of $\rho_{AB}$ and $W_{\beta_j(x,s)}(\mathbf{R}_B)$
is the Wigner function of the coherent state $|\beta_j(x,s)\rangle$. 
With a lengthly but straightforward calculation we found that the Wigner function of $\rho_{A_j}(x,s)$ 
is also Gaussian with covariance matrix
\begin{eqnarray}
V_{A_j(x,s)} = \left[ \frac{(1-\eta)N}{1+\eta N} + \frac{1}{2} \right]
\left(
\begin{array}{cc}
1 & 0 \\
0 & 1
\end{array}
\right) \, .
\end{eqnarray}
From $V_{A_j(x,s)}$ we can compute the von Neumann entropy of the conditional states $\rho_{A_j}(x,s)$, which
is $S(\rho_{A_j}(x,s)) = g[(1-\eta)N']$ with $N' = N/(1+\eta N)$.

By applying the same reasoning we compute the covariance matrix of Eve's conditional states $\rho_{E_j}(x,s)$:
\begin{eqnarray}
V_{E_j(x,s)} = \left[ \frac{(1-\eta)N}{1+\eta N} + \frac{1}{2} \right]
\left(
\begin{array}{cc}
1 & 0 \\
0 & 1
\end{array}
\right) \, .
\end{eqnarray}
We also compute the mean $\bar{\mathbf{R}}_j = (\bar{q}_{E_j},\bar{p}_{E_j})$ and obtain
\begin{eqnarray}
\bar{q}_{E_j(x,s)} & = & \frac{N \sqrt{\eta (1-\eta)}}{1+\eta N} \, \frac{\mathsf{Re}[\beta_j(x,s)]}{\sqrt{2}} \\
\bar{p}_{E_j(x,s)} & = & \frac{N \sqrt{\eta (1-\eta)}}{1+\eta N} \, \frac{\mathsf{Im}[\beta_j(x,s)]}{\sqrt{2}} \, .
\end{eqnarray}
Notice that the mean is also a function of the mode label $j$ through the 
amplitude $\beta_j(x,s)$. 
We remark that Alice's and Eve's conditional states 
have the same covariance matrix but different mean.

\subsection{Calculations for the security proof}\label{ss3}

From the form of the conditional state 
$\rho_E^n(x,s) = \bigotimes_{j=1}^n \rho_{E_j}^n(x,s)$ we can compute
\begin{equation}
\mathbb{E}_s[\rho_E^n(x,s)] = \rho_E^{\otimes n}
\end{equation}
(notice that, for how Bob's measurement has been defined, the expectation
value over $s$ equals the expectation value over $x$).
$\rho_{E}$ is a Gaussian state with zero mean. 
Its covariance matrix can be obtained directly from (\ref{BE-CM}) and reads
\begin{align}
V_{E} & = \frac{1}{2} \left(
\begin{array}{cc}
C(1-\eta)+\eta           & 0 \\
0                        & C(1-\eta)+\eta
\end{array}
\right) \\
& = \left[ (1-\eta)N + \frac{1}{2} \right]
\left(
\begin{array}{cc}
1           & 0 \\
0           & 1
\end{array}
\right) \, .
\end{align}
That is, $\rho_{E}$ is a thermal state with $(1-\eta)N$ mean photons, whose entropy is
$S(\rho_{(1-\eta)N}) = g[(1-\eta)N]$. We then obtain
\begin{equation}
2^{-n [ g[(1-\eta)N] + \delta]} \, \mathbb{P}_\rho^n \leq \mathbb{P}_\rho^n \, \rho_E^{\otimes n} \mathbb{P}_\rho^n 
\leq 2^{-n [ g[(1-\eta)N] - \delta]} \, \mathbb{P}_\rho^n \, .
\end{equation}
The spectral decomposition of $\rho_E$ is as in Eq.\ (\ref{spectre1}).
Similarly, the operator $\rho_E \otimes \rho_E$ is identical to its homologous
analyzed for the direct reconciliation protocol, with spectral decomposition 
given in Eq.\ (\ref{spectre2}).

We now consider the operator
$\rho_{2E} = \mathbb{E}_s[ \rho_{E_j}(x,s) \otimes \rho_{E_j}(x,s) ]$. 
Using the results of Sec.\ \ref{ss2} we found that $\rho_{2E}$ is a Gaussian
state with zero mean and covariance matrix
\begin{widetext}
\begin{eqnarray}
V_{2E} = \left( \begin{array}{cccc}
(1-\eta)N+\frac{1}{2} & 0                     & \eta(1-\eta)N N'      & 0 \\
0                     & (1-\eta)N+\frac{1}{2} & 0                     & \eta(1-\eta)N N' \\
\eta(1-\eta)N N'      & 0                     & (1-\eta)N+\frac{1}{2} & 0 \\
0                     & \eta(1-\eta)N N'      & 0                     & (1-\eta)N+\frac{1}{2}
\end{array} \right) \, ,
\end{eqnarray}
\end{widetext}
with $N' = N/(1+\eta N)$.
From the covariance matrix $V_{2E}$ we compute the von Neumann entropy
$S(\rho_{2E}) = g[(1-\eta)N'] + g[(1-\eta)N'']$, with $N'' = N(1+2\eta N)/(1+\eta N)$.
Finally, its spectral decomposition is
\begin{align}
\rho_{2E} & = \frac{1}{(1-\eta)^2N' N''} 
\sum_{t,m=0}^\infty 
\left( \frac{(1-\eta)N'}{(1-\eta)N'+1} \right)^t \nonumber \\
& \times \left( \frac{(1-\eta)N''}{(1-\eta)N''+1} \right)^m
|\psi_{t,m} \rangle \langle \psi_{t,m}| \, ,
\end{align}
with
\begin{align}
|\psi_{t,m} \rangle & = 2^{\frac{-t-m}{2}}
\sum_{j=0}^t \sum_{k=0}^m { t \choose j} { m \choose k} (-1)^k \sqrt{(t+m-j-k)!} \nonumber \\
& \times \sqrt{(j+k)!} \, |t+m-j-k\rangle |j+k\rangle \, .
\end{align}
Notice that $|\psi_{t,m} \rangle$ is a state with exactly $\ell = t + m$
photons. It follows that $\rho_{2E}$ commutes with $\rho_E \otimes \rho_E$ (see Eq.\ (\ref{spectre2})).

\subsection{Active attack}

\begin{figure}
\centering
\includegraphics[width=0.4\textwidth]{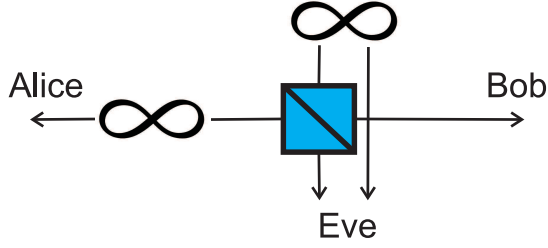}
\caption{
A scheme for an active Gaussian attack. The beam splitter mixes Alice's mode with one mode
from an entangled pair (denoted by the symbol `$\infty$'). 
The eavesdropper Eve obtains both the modes of the two-mode state.}
\label{fig:active}
\end{figure}

An active Gaussian attack from the eavesdropper can be modeled as a beam-splitter
that mixes the mode from Alice with a mode from a two-mode entangled state.
As shown in Fig.\ \ref{fig:active}, the eavesdropper Eve obtains both the modes of the
two-mode entangled state. In this setting, if Alice's two-mode entangled state has
$N$ mean photons per mode, and Eve's two-mode entangled state has $N_T$ mean photons per
mode, then the joint four-mode Gaussian state of Alice, Bob
and Eve has covariance matrix:
\begin{widetext}
{\footnotesize 
\begin{align}
& V_{ABEE'} = \nonumber \\
&\frac{1}{2} \left( \begin{array}{cccccccc}
C              & 0               & S\sqrt{\eta}               & 0                          & S\sqrt{1-\eta}             & 0                          & 0                 & 0 \\ 
0              & C               & 0                          & -S\sqrt{\eta}              & 0                          & -S\sqrt{1-\eta}            & 0                 & 0 \\
S\sqrt{\eta}   & 0               & C_T(1-\eta)+C\eta          & 0                          & (C-C_T)\sqrt{\eta(1-\eta)} & 0                          & -S_T\sqrt{1-\eta} & 0 \\
0              & -S\sqrt{\eta}   & 0                          & C_T(1-\eta)+C\eta          & 0                          & (C-C_T)\sqrt{\eta(1-\eta)} & 0                 & S_T\sqrt{1-\eta} \\
S\sqrt{1-\eta} & 0               & (C-C_T)\sqrt{\eta(1-\eta)} & 0                          & C(1-\eta)+C_T\eta          & 0                          & S_T\eta           & 0 \\
0              & -S\sqrt{1-\eta} & 0                          & (C-C_T)\sqrt{\eta(1-\eta)} & 0                          & C(1-\eta)+C_T\eta          & 0                 & -S_T\eta \\
0              & 0               & -S_T\sqrt{1-\eta}          & 0                          & S_T\eta                    & 0                          & C_T               & 0 \\
0              & 0               & 0                          & S_T\sqrt{1-\eta}           & 0                          & -S_T\sqrt{\eta}            & 0                 & C_T
\end{array}\right) \, ,
\end{align}
}
\end{widetext}
where $C = 2N+1$, $S = 2\sqrt{N(N+1)}$, and $C_T = 2N_T+1$, $S_T = 2\sqrt{N_T(N_T+1)}$.
We can use this covariance matrix instead of (\ref{CM-lossy}) and repeat the calculations
done in subsections \ref{ss1}-\ref{ss3} for the reverse reconciliation protocol.
We obtain
\begin{equation}
\chi_\mathrm{rr} = g(N) - g[(1-\eta)\tilde{N}] \, ,
\end{equation}
with $\tilde{N} = N(1+N_T)/[1+N_T-(N-N_T)\eta]$, and, for $N \gg 1,N_T$,
\begin{equation}
k_\mathrm{rr} = g(N_T) + 2g[(1-\eta)N+\eta N_T] - g[(1-\eta)\tilde{N}] - g[\hat{N}] \, ,
\end{equation}
with $\hat{N} = 2(1-\eta)N + \frac{(1-\eta)+N_T(2\eta^2-1)}{\eta}$.
Finally, in the limit $N \to \infty$ we obtain
\begin{equation}
r_\mathrm{rr} = \chi_\mathrm{rr} - k_\mathrm{rr} = 1 + \log{\left(\frac{1}{1-\eta}\right)} - g(N_T) \, .
\end{equation}

\section{Fluctuations of the mean photon number $\Delta \langle \ell \rangle$}

Let us consider the distribution $\pi$, with
\begin{equation}\label{tpi}
\pi_\ell = \frac{1}{(1-\eta)N +1} \left( \frac{(1-\eta)N}{(1-\eta)N+1} \right)^\ell (\ell+1) \, ,
\end{equation}
and a $\delta$-typical type $\tilde\pi$. 
The empirical entropy given by $\tilde\pi$ is
\begin{equation}
S = - \sum_{\ell=0}^\infty \tilde\pi_\ell \log{\pi_\ell} \, .
\end{equation}
For $\delta$-typical type we have small fluctuation of $S$ around its average, that is,
\begin{equation}
\Delta S = - \sum_{\ell=0}^\infty \tilde\pi_\ell \log{\pi_\ell} + \sum_{\ell=0}^\infty \pi_\ell \log{\pi_\ell} \in [-c\delta,c\delta] \, .
\end{equation}

From (\ref{tpi}) we obtain
\begin{align}
S & = \log{[(1-\eta)N +1]} \nonumber \\
& - \log{\left( \frac{(1-\eta)N}{(1-\eta)N+1} \right)} \langle \ell \rangle_{\tilde\pi}  
- \langle \log{(\ell+1)}\rangle_{\tilde\pi} \, ,
\end{align}
which yields
\begin{equation}
\Delta S = - \log{\left( \frac{(1-\eta)N}{(1-\eta)N+1} \right)} \Delta \langle \ell \rangle  
- \Delta \langle \log{(\ell+1)}\rangle \, .
\end{equation}
For $N$ large enough we have
\begin{equation}
\Delta S \simeq \log{e} \left[ \frac{\Delta \langle \ell \rangle}{(1-\eta)N}  
- \frac{\Delta \langle \ell \rangle}{(1-\eta)N+1} \right] \, ,
\end{equation}
where we have used the fact that $\langle \ell \rangle_{\tilde\pi}$ fluctuates about
$(1-\eta)N$. Finally we obtain
\begin{equation}
\frac{\log{e} \, \Delta \langle \ell \rangle}{(1-\eta)N } \simeq [ (1-\eta)N + 1 ] \Delta S \, .
\end{equation}
Since for $N$ large enough $\beta = \frac{\log{e}}{(1-\eta)N}$, we have 
\begin{equation}
\beta \Delta \langle \ell \rangle \simeq [ (1-\eta)N + 1 ] \Delta S \, .
\end{equation}


\begin{thebibliography}{99}
\bibitem{NOTA_tdist}
We recall that given a hermitian operator $X$, $\| X \|_1 := \mathrm{Tr}|X|$.
\bibitem{NOTA_POVM}
We recall that the POVM elements have the properties $ \Lambda_y \geq 0$ 
and $\sum_y \Lambda_y = \mathbb{I}$.
\bibitem{NOTA_acc}
We recall that the accessible information is the maximum mutual
information that can be achieved by local measurements, i.e.,
$$
I_\mathrm{acc}(X;E)_{\rho} = \max_{\Lambda} I(X;Y) \, ,
$$
where the maximum is over measurements $\Lambda$ applied on Eve's
system and $I(X;Y) = H(X) + H(Y) - H(XY)$ is the mutual information
between the message and the output of Eve's measurement.
\bibitem{NOTA_Pinsker}
Given the joint probability distribution $p_{XY}$ and the marginals
$p_X$, $p_Y$, Pinsker inequalities states that
\begin{equation}
D(p_{XY}, p_X p_Y) \leq \sqrt{ 2\ln{(2)} \, I(X;Y) } \, .
\end{equation}
\bibitem{NOTA_superb} 
To achieve this rate Alice may need to make collective measurements over super-blocks of $n'n$ modes.
\end{thebibliography}
\end{document}